\documentclass[preprint,prb,showpacs]{revtex4}
\usepackage[english]{babel}
\usepackage[latin1]{inputenc}
\usepackage[dvips]{graphicx}
\usepackage{amsmath}
\usepackage{amssymb}
\usepackage{epsfig}
\usepackage{array}

\begin{document}

\title{Diffusion in a generalized Rubinstein-Duke model of electrophoresis
with kinematic disorder.}
\author{Richard D. Willmann}
\email{r.willmann@fz-juelich.de}
\author{Gunter M. Sch\"utz}
\affiliation{Institut f\"ur Festk\"orperforschung, Forschungszentrum J\"ulich,
52425 J\"ulich, Germany}
\author{Kavita Jain}
\affiliation{Department of Theoretical Physics, Tata Institute of
Fundamental Research, Homi Bhabha Road, Mumbai 400 005, India}

\date{\today}

\begin{abstract}
Using a generalized Rubinstein-Duke model we prove rigorously that kinematic disorder 
leaves the prediction of standard reptation theory for the scaling of the diffusion 
constant in the limit for long polymer chains $D \propto L^{-2}$ unaffected. Based on 
an analytical calculation as well as Monte Carlo simulations we predict kinematic 
disorder to affect the center of mass diffusion constant of an entangled polymer in 
the limit for long chains by the same factor as single particle diffusion in a random 
barrier model.
\end{abstract}


\maketitle

The derivation of large scale properties of polymer systems such as viscosity and 
diffusion constant out of microscopic models is among the basic problems of polymer 
science. For entangled polymers in gels or melts, de Gennes \cite{deGennes} predicted 
scaling laws for the dependence on the polymer length $L$ for viscosity ($\eta \propto 
L^3$) and diffusion constant ($D \propto L^{-2}$). These scaling laws are assumed to 
be valid for the limiting case of polymer length going to infinity. In experiments, 
the apparent scaling laws $D \propto L^{-2.4}$ and $\eta \propto L^{3.4}$ are found
\cite{Lodge1,Lodge2,Ferry}. However, the experimental findings do not contradict the 
predictions of reptation, as a crossover due to decreasing finite size effects with 
increasing chain length cannot be ruled out \cite{Lodge2}. Presumably, contour length 
fluctuations (CLF) are one cause for the deviating scaling exponent for finite chain 
length \cite{Doi,Enrico,Matthias}. In the framework of the repton model introduced by 
Rubinstein \cite{Rub} and Duke \cite{Duk,Duke2} (further on to be called RD model), 
which incorporates CLF, it is possible to calculate viscosity, diffusion coefficient 
and other quantities of interest. Good agreement both with theoretical and experimental 
results is found \cite{Enrico,Matthias,Exp1,Exp2}. Using the RD model with periodic 
boundary conditions Kooiman and van Leeuwen \cite{Koi} analytically calculated the 
proportionality constant $c$ for the diffusion constant in the limit for infinite 
chain length: $\lim_{L \to \infty} DL^2=c$. They found $c=w/(2d+1)$, $d$ being the 
dimensionality of the entanglement network and $w$ a model constant defining the time 
scale. Building upon this result, Pr\"ahofer and Spohn \cite{Prae} rigorously derived 
the leading order term in $1/L$ for the diffusion constant in the RD model and 
furthermore proposed a scaling for the finite size effects: $DL^2-W/(2d+1) \propto 
L^{- \beta}$, where $1/2 \leq \beta \leq 1$. All results mentioned above were obtained 
under the assumption that the entanglement network which is topologically restricting 
the polymer is regular and static. Real entanglement networks such as gels or melts 
are disordered. The effects of a disordered environment can be manifold:

\begin{itemize}
\item Spatial variations of the mobility of the 'defects' of stored length.
\item Locally fluctuating potential energy due to interactions between chain and 
environment.
\item Entropically favorable regions of low entanglement density.
\item Relaxation of the environment (constraint release).
\end{itemize}

Numerical investigations \cite{Mut} showed that entropically favorable regions can for 
short chains substantially lower the diffusion constant by the creation of 'entropic 
traps'. So far, conclusive investigations of diffusing polymers long enough to span 
several such traps are missing. Constraint release is considered to be of minor 
importance in gels but needs to be self consistently taken into account in melts 
\cite{McLeish}.
Sch\"afer, Baumg\"artner and Ebert \cite{Ebert} numerically investigated the effect 
of kinematic disorder, i.e., disorder which affects the mobility of the chain segments 
while leaving the equilibrium distribution of chain configurations unaltered. Their 
investigation shows that reptation prevails in presence of kinematic disorder. However, 
due to being based on MC simulations and thus relatively short chains no quantitative 
results could be obtained about the modifications the disorder would cause to the 
reptation prediction for the diffusion constant in the long chain limit. As reptation 
is shown to prevail, the scaling $\lim_{L \to \infty} DL^2=c$ must remain valid. The 
aim of this article is to calculate the coefficient $c$ for a polymer diffusing in a 
disordered environment exhibiting kinematic disorder. This means that we are interested 
in the behavior of the chain in the long chain length limit originally envisaged by 
standard reptation theory. Some of the results presented here in detail were briefly 
reported in an earlier communication \cite{Richard}.\\
The article is divided into six chapters: In chapter I the definition of the model is 
given and the master equation for the chain dynamics is presented in terms of the 
quantum Hamiltonian formalism \cite{Habil}. This master equation yields the stationary 
distribution of chain conformations as shown in chapter II. The model with periodic 
boundary conditions is analyzed in chapter III. Adapting the strategy in Ref. 
\onlinecite{Koi} for obtaining the diffusion constant for the periodic system yields a 
lower bound on the diffusion constant for the original system. Chapter IV is devoted 
to deriving an upper bound to the diffusion constant by generalizing a variational 
approach used in Ref. \onlinecite{Prae} for the ordered system. In chapter V 
Monte Carlo simulations of the polymer diffusion in kinematically disordered 
environments are presented. The last chapter addresses the dynamics of internal chain
segments.

\section{A lattice model for reptation with kinematic disorder}

\begin{figure}[h!]
\begin{center}
\includegraphics[scale=1.3]{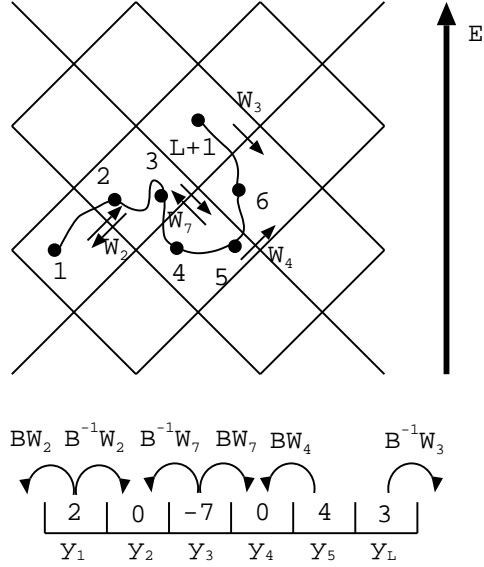}
\caption{Two dimensional representation of a kinematically disordered lattice with
a reptating polymer chain and the lattice gas mapping. Arrows indicate possible
moves of the reptons. $B=\exp(E/2)$.}
\label{model}
\end{center}
\end{figure}

The RD model is a discretized model for reptation, i.e., it describes the dynamics of 
a polymer in an entanglement network. This network is assumed to be regular and static. 
In 3 dimensions it has the shape of a cubic lattice, where the polymer is forbidden to 
cut through the edges of the cubes. The faces of the cubes can by penetrated by the 
chain (see figure \ref{model}). The polymer itself is assumed to consist of $L+1$ 
'reptons', i.e., segments of about the entanglement length, which equals the lattice 
constant of the cubic lattice.

The model defines dynamical rules for the reptons' motion. The dynamics on a smaller 
scale, i.e. high frequency Rouse modes, is averaged out. In this sense, the RD model 
is an effective model. The dynamical rules for the reptons are:

\begin{itemize}
\item[a)] Each cell occupied by the chain must contain at least one repton to ensure 
connectivity of the chain, since the length represented by a repton equals the lattice 
constant of the cells. The sequence of occupied pores corresponds to the tube as in 
standard reptation theory. 
\item[b)] End reptons can move to adjacent cells provided rule a) is not violated.
\item[c)] Interior reptons can move to cells occupied by the neighboring reptons if 
allowed by rule a). This ensures the dynamics to be reptation, as any movements in 
the interior requires multiple reptons to occupy the same cell, which means that there 
is some excess 'stored length' available. This corresponds to the motion of 'defects' 
in the original de Gennes reptation model.
\end{itemize}

All moves are assumed to be thermally activated. In our model the cubic lattice 
representing the entanglement network gets kinematically disordered by assigning an 
individual energy barrier to every boundary between adjacent cells. These barriers 
have to be overcome by a crossing repton. Note that we do not assign different energies 
to the reptons when residing in the cells. This choice of quenched disorder guarantees 
that only the mobility of the reptons, but not the equilibrium configurations are 
affected, as will be seen below. Let there be $\sigma \in N$ different possible hopping 
rates $W_{\alpha}$, which are distributed randomly and occur with a probability 
$f(W_{\alpha})$ throughout the lattice. We demand that the disorder averages 
$\langle 1/W \rangle=\sum_{\alpha=1}^{\sigma} f(W_{\alpha})/W_{\alpha}$ and 
$\langle 1/W^2 \rangle$ (the mean hopping time and its second moment) are finite. 

The RD model is used to describe the dynamics of an entangled polymer chain under the 
influence of an external electric field. A common example is DNA under electrophoresis 
conditions where the reptons carry a charge each and develop a drift velocity along 
the direction of an applied electric field. Let the field be oriented along the 
$(111)$ diagonal of the cubes. We denote the (dimensionless) energy gain of a repton 
when moving from one cell to another along the direction of the field as $E$. By local 
detailed balance, moves across a cell boundary with assigned hopping rate $W_{\alpha}$ 
along the field happen with rate $W_{\alpha} \exp(E/2)$, those in the opposite 
direction with rate $W_{\alpha} \exp(-E/2)$ \cite{Duk}. By projecting the reptons' 
positions on the direction of the field, their relative coordinates can be denoted as 
a 1-d lattice gas with $L$ sites by the following prescription:

\begin{itemize}
\item If the projected link between adjacent reptons $i$ and $i+1$ is oriented along 
(against) the field direction across a cell boundary with assigned rate $W_{\alpha}$, 
site $i$ of the lattice gas is assigned the value $y_i=\alpha$ ($y_i=- \alpha$). We 
interpret values $\pm  \alpha $ as particles.
\item If adjacent reptons $i$ and $i+1$ occupy the same cell, i.e., their projected 
positions coincide, the link is represented in the lattice gas as $y_i=0$ at site $i$. 
We interpret a site, which is assigned a 0 as being unoccupied. 
\end{itemize}

\begin{figure}[h!]
\begin{center}
\includegraphics[scale=0.3]{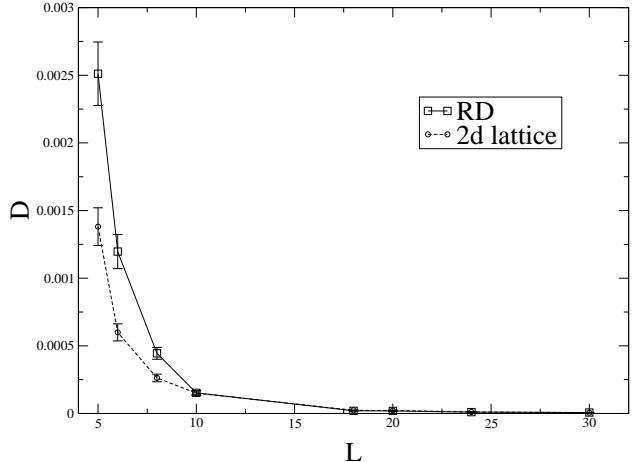}
\caption{Comparison between the diffusion coefficients for simulations of the 
disordered RD and two dimensional lattice model.}
\label{disorder}
\end{center}
\end{figure}

Thus the $L+1$ coordinates of the reptons in direction of the field are translated 
into the equivalent set of the relative coordinates manifest as the assignment of 
$\alpha , - \alpha  $ or 0 to the $L$ sites of the lattice gas plus the center of mass 
coordinate's component in field direction. The dynamics of the reptons translates 
into the lattice gas picture as follows:
In the bulk particles of sort '$\pm \alpha$' hop to the left with rate $W_{\alpha} 
\exp(\pm E/2)$ and to the right with rate $W_{\alpha} \exp(\mp E/2)$, where each site 
can be occupied by at most one particle. The end dynamics in the lattice gas picture 
needs some care: Assuming $y_1$ ($y_L$) to be non zero, the only possible move is the 
retraction of the end repton to the cell occupied by it's neighbor (rule a)). This 
retraction, being a particle annihilation event in the lattice gas picture, happens 
with the same rate as the respective move in the bulk. Assuming $y_1$ ($y_L$) to be 
zero, the end repton can, according to rule b), move to any of the $2d$ adjacent cells. 
For half of these the move leads to links being along the field direction, the other 
half against it. The probability of the chosen move leading to the crossing of a cell 
boundary with rate $W_{\alpha}$ being assigned to it is $f(W_{\alpha})$. Thus the move 
of the repton, being a particle creation event in the lattice gas picture leads to 
$y_1$ ($y_L$) changing from '0' to '$\pm \alpha$' with rate 
$\exp(\mp E/2)f(W_{\alpha})W_{\alpha}d$ ($\exp(\pm E/2)f(W_{\alpha})W_{\alpha}d$). 
This choice of boundary dynamics is on average correct, but neglects the actual local 
structure of the network. In order to verify that the model is correct on long time 
scales (as relevant for the CMS diffusion) we performed Monte Carlo simulations 
comparing the diffusion constant of our projected model with $d=2$ at different 
disorder distributions with a polymer chain moving according to the dynamics of the 
repton model in a two dimensional lattice with random but fixed hopping rates assigned 
to each cell boundary. It turns out that although the results differ for small chains 
($L<10$) they coincide within the statistical errors for longer chains. This 
legitimates our choice of boundary dynamics for investigating the behavior of long 
chains. Figure \ref{disorder} shows an example where a disorder distribution with $<W>=0.505$ was 
chosen. \\

Every move of a particle in the lattice gas leads to a change in the component of the 
center of mass coordinate of the repton chain along the field direction.

\begin{itemize}
\item Particle type '$\alpha$' moving to the right (left) decreases (increases) $x$ by 
$1/(L+1)$, as this is equivalent to a repton moving downward (upward). As there are 
$L+1$ reptons each contributes $1/(L+1)$ to the center of mass position.
\item Particle type '$-\alpha$' moving to the right (left) increases (decreases) $x$ 
by $1/(L+1)$.
\end{itemize}

In the subsequent chapters we will calculate the drift velocity $v$ of the center of 
mass coordinate in presence of an applied electric field and by employing the Einstein 
relation the zero field diffusion constant. When calculating $v$ we restrict ourselves to 
the linear response regime, ignoring higher order field dependences.
In order to calculate $v$ only the change 
in the center of mass coordinate along the field $x$ has to be known, but not the 
absolute value itself. Whereas the latter cannot be known from the lattice gas, the 
former is given by the difference of the  currents of positive particles $j^+$ and that 
of the negative ones $j^-$: $v=j^--j^+$. As the choice of the field direction is 
arbitrary, by use of the Einstein relation the model allows for computing the zero 
field diffusion constant along a distinct direction in $d$-dimensional space. As zero 
field diffusion is isotropic this immediately yields the $d$-dimensional diffusion 
constant. This is due to distinguishing between particles of types $\alpha$ and 
$- \alpha$ in contrast to the original projected repton model used by Rubinstein, which 
allows only for the computation of the curvilinear diffusion constant along the contour 
of the chain. In the Rubinstein model additional assumptions are necessary to relate 
the curvilinear to the $d$-dimensional diffusion constant. Our model as well as the RD 
model allows for computation of the latter quantity within the model.

\section{Quantum Hamiltonian and stationary state}

The model introduced in the previous chapter describes a Markov process and thus the 
dynamics can be written in form of a master equation. For convenience we will use the 
quantum Hamiltonian formalism \cite{Habil} to write down the master equation at zero 
electric field and solve for the stationary state. \\
At each site of the lattice gas with length $L$, $2 \sigma +1$ values for $y_i$ are 
possible. Therefore, the state space $X$ has the dimension $(2 \sigma +1)^L$. Every 
$\eta \in X$ is assigned a vector $| \eta >$ and a transposed vector $< \eta |$. These 
vectors constitute a basis for the space of system configurations $X$. A probability 
distribution $ P(\eta) \equiv P_{\eta}$ can thus be written as a probability vector: 

\begin{equation}
|P>= \sum_{ \eta \in X} P(\eta) | \eta >.
\end{equation}

Let a summation vector $<s|$ be defined as $ <s|= \sum_{ \eta \in X} < \eta | $. 
Normalization of the probability vector is given if $<s|P>=1$. The generator for the 
dynamics of the system is the matrix $ H $: The off-diagonal elements of the matrix 
contain the (negative) transition rates $ t(\eta,\eta ') $ from a state $ \eta '$ to 
$ \eta $:

\begin{equation}
< \eta |H| \eta ' > = H_{\eta , \eta '} = -t(\eta,\eta ').
\end{equation}

The diagonal elements $ H_{\eta , \eta} $ contain the sum of all outgoing rates :

\begin{equation}
< \eta |H| \eta > = H_{\eta , \eta } = \sum_{\eta ' \neq \eta} t(\eta ',\eta). 
\end{equation}

It is easily checked that conservation of probability, i.e. $ <s|H = 0 $ in the 
language of the quantum Hamiltonian formalism, is fulfilled by $ H $. The master 
equation
 
\begin{equation}
\frac{d}{dt}P_{\eta}(t)=\sum_{\eta \neq \eta', \eta \in X} \bigg ( t( \eta, \eta')P_{ 
\eta'}(t)-t( \eta', \eta) P_{ \eta}(t) \bigg )
\end{equation}

describing the Markovian dynamics of the system can be written as:

\begin{equation}
 \frac{d}{dt} |P(t)> = -H|P(t)>.
\end{equation}

The stationary state $ |P^*> $ is thus characterized by the equation

\begin{equation}
H|P^*> = 0.
\end{equation}

In this formalism the expectation value $<F(t)>= \sum_{\eta} F(\eta)P(\eta,t)$ of an 
operator $F$ is written as follows \cite{Habil}: The operator $F: X \rightarrow X$ is 
represented by a diagonal matrix $F=\sum_{\eta} F(\eta) |\eta><\eta|$. Then 

\begin{equation}
\begin{split}
<F(t)> & =<s|F e^{-Ht}|P(0)> \\
& = <s|e^{Ht}F e^{-Ht}|P(0)>. \\
\end{split}
\end{equation}

Let the time-dependent operator $F(t)$ for $t>0$ be defined as 

\begin{equation}
F(t)=e^{Ht}F e^{-Ht}.
\end{equation}

$<F(t)>$ is an expectation value that it is not only averaged over possible realizations
 of the process but also over the initial states according to $P(0)$. In the following 
it is assumed that $|P(0)>=|P^*>$, so that 

\begin{equation}
<F(t)> = <s|F(t)|P^*>.
\end{equation}

For our model, we choose a tensor product basis as follows: Let at each site of the 
lattice gas the unit vector $e_1$ denote $y_i=0$, $e_{2 \alpha}$ denote $y_i= \alpha$ 
and $e_{2 \alpha+1}$ denote $y_i= - \alpha$. A state vector for a state 
$\eta=(1,-3,..,5,-2)$ for example can then be written as 
$| \eta >= e_2 \otimes e_7 \otimes ... \otimes e_{10} \otimes e_5)$. Using this basis, 
the following operator creates a particle of type $\alpha$ at site $i$, provided it was 
previously unoccupied:

\begin{equation}
a^{\dagger}_{\alpha,1}(i)= \underbrace{ \mathbf{1} \otimes .. \otimes}_{i-1} E_{(2 
\alpha,1)} \underbrace{\otimes .. \otimes \mathbf{1}}_{L-i} = E_{(2 \alpha,1)}(i),
\end{equation}

where $E_{(2 \alpha,1)}$ is the matrix with a single entry 1 at row $2 \alpha$ and 
column 1. Similarly, the operator $a^{\dagger}_{\alpha,-1}(i)=E_{(2 \alpha+1,1)}(i)$ 
creates a particle of type $- \alpha$, if possible. The corresponding annihilation 
operators at site $i$ are $a_{\alpha,1}(i)=E_{(1,2 \alpha)}(i)$ and 
$a_{\alpha,-1}(i)=E_{(1,2 \alpha+1)}(i)$. To formulate the diagonal part of the quantum 
Hamiltonian the matrices $u(i)=E_{(1 , 1) }(i) $, $ v_{\alpha,1}(i)=E_{ (2\alpha , 
2\alpha)}(i) $ and $ v_{\alpha,-1}(i)=E_{( 2\alpha+1 , 2\alpha+1)}(i) $ are employed. 
Thus, the quantum Hamiltonian for the model defined in the previous chapter at zero 
field reads:

\begin{eqnarray}
 H_{open} & = & \sum_{i=1}^{L-1} \sum_{s= \pm 1} \sum_{\alpha=1}^{\sigma} W_\alpha  
\big [ -a_{\alpha,s}^{ \dagger}(i+1) a_{\alpha,s}(i) - a_{\alpha,s}^{ \dagger}(i)
a_{\alpha,s}(i+1) \notag \\ 
& & + u(i+1) v_{\alpha,s}(i) +  u(i) v_{\alpha,s}(i+1) \big ] \notag \\ 
& & + \sum_{s= \pm 1} \sum_{\alpha=1}^{\sigma} \big [ W_\alpha  d  f(W_\alpha)
(-a_{\alpha,s}^{ \dagger}(1) +  u(1)) + W_\alpha (-a_{\alpha,s}(1) +v_{\alpha,s}(1)) 
\big ] \notag \\
& & + \sum_{s= \pm 1} \sum_{\alpha=1}^{\sigma} \big [ W_\alpha  d   f(W_\alpha) 
(- a_{\alpha,s}^{ \dagger}(L)+ u(L)) \notag \\
& & + W_\alpha (-a_{\alpha,s}(L) + v_{\alpha,s}(L)) \big ]
\end{eqnarray}

Plugging a product measure ansatz

\begin{equation}
|P_{open}^*(0)> = \begin{pmatrix} 1 \\ p_{1,1}(1) \\ p_{1,-1}(1) \\ p_{2,1}(1) \\ 
\vdots \end{pmatrix} \otimes \ldots \otimes \begin{pmatrix} 1 \\ p_{1,1}(L) \\ p_{1,-1}
(L) \\ p_{2,1}(L) \\ \vdots \end{pmatrix}  \frac{1}{ \prod_{i=1}^L (1+ 
\sum_{\alpha=1}^{\sigma} (p_{\alpha,1}(i)+p_{\alpha,-1}(i)))}
\end{equation}

into $H_{open}|P_{open}^*(0)>=0$ leads to a very simple set of equations for the 
probabilities $p_{\alpha,\pm 1}$ and one finds

\begin{equation}
|P_{open}^*(0)> =  \begin{pmatrix} 1 \\ d  f(W_1) \\ d  f(W_1)  \\ \vdots \\ d  
f(W_{\sigma})  \\ d  f(W_{\sigma}) \end{pmatrix}^{ \otimes L }  \frac{1}{(2  d +1 )^L}.
\end{equation}

The geometrical equilibrium conformation of the chain depends on the probability of 
occurrence for links between reptons along or against the field, irrespective of the 
assigned hopping rates of possibly crossed cell boundaries. This means we have to 
consider the overall probability for particles of positive sign at a site, which is 
$\sum_{ \alpha =1}^{ \sigma} f(W_{ \alpha})d / (2d+1)=d/(2d+1)$, or negative sign, 
being also $d/(2d+1)$, respectively. These are equal to the probabilities found for 
the disorder free RD-model \cite{Diplom} which shows that the chosen kind of disorder 
is indeed kinematic disorder, as it influences only the mobility but not the equilibrium
 configurations of the chain. \\
Note that it is not possible to compute the stationary state for the model with a 
non-zero field at arbitrary $L$. Therefore the Einstein relation $1/L \enspace d/dE 
(<v(E)>)=D$ is not a feasible way of straightforwardly computing the diffusion constant.
 Extra input is needed to overcome this problem.

\section{Model with periodic boundary conditions}

In this section we adapt a calculation due to Kooiman and van Leeuwen \cite{Koi} for the
 original RD-model with periodic boundary conditions to the case with kinematic disorder.
 The quantum Hamiltonian for the periodic system at zero field reads

\begin{eqnarray}
H_{per} & = & \sum_{s= \pm 1} \sum_{\alpha=1}^{\sigma}  \sum_{i=1}^{L} W_\alpha  \big 
[ -a_{\alpha,s}^{ \dagger}(i+1) a_{\alpha,s}(i) -  a_{\alpha,s}^{ \dagger}(i)
a_{\alpha,s}(i+1) \notag \\ 
& & + u(i+1) v_{\alpha,s}(i) +  u(i) v_{\alpha,s}(i+1) \big ].
\end{eqnarray}

Plugging $|P_{open}^*(0)>$ as for the open system into the equation 
$H_{per}|P_{open}^*(0)>=0$ for the periodic system shows that at zero electric field 
the stationary state of the open system is also stationary with respect to the dynamics 
of a corresponding periodic system. In a periodic system, the phase space is non ergodic,
 as neither the order nor the number of occurring particles on such a ring can be 
changed. Therefore, every connected subset of the phase space ('channel') has its own 
stationary state. We can calculate the stationary state for the periodic system in 
presence of a field $E$ by mapping the system to a disordered zero range process, as 
introduced by Benjamini, Ferrari and Landim \cite{Ben}. This means that the Einstein 
relation can be employed to obtain the diffusion constant in the periodic case. Here, 
the definition of the drift velocity of the center of mass and the corresponding 
diffusion constant are induced from the system with open boundary conditions: 
$v=j^--j^+$.\\
Instead of characterizing the system by the spins on the lattice gas 
$\mathbf{y}=(y_1,..,y_L)$ it can equivalently be characterized by the sets 
$\mathbf{s}=(s_1,..,s_L)$ containing the signs of the non zero $y_i$, 
$\mathbf{w}=(w_1,..,w_L)$ containing the absolute values of these $y_i$ and 
$\mathbf{n}=(n_1,..,n_L)$ where $n_i$ amounts to the number of '0' between $s_i$ and 
$s_{i+1}$ on the lattice gas. Thus every lattice gas configuration on a ring can be 
translated into a configuration on a lattice of length $M$ in the zero range picture: 
Site $i$ carries $n_i$ zero range particles and is separated from site $i+1$ by a bond 
characterized by $s_{i+1}$ and $w_{i+1}$. The total number $K$ of zero range particles 
must equal the total number $L-M$ of '0' in the lattice gas: $\sum_{i=1}^M n_i=K=L-M$. 
The dynamics of the lattice gas picture translates into the zero range picture as 
follows: The configuration $(..,n_j,n_{j+1},..)$ changes to $(..,n_j-1,n_{j+1}+1,..)$ 
with rate $h_{j+1}^{-1}w_{j+1}$ and to $(..,n_j+1,n_{j+1}-1,..)$ with rate 
$h_{j+1}w_{j+1}$, where $h_j=\exp(-Es_j/2)$. This means that the random hopping rates 
as well as the $s_j$ are not assigned to individual particles, but to bonds between 
sites in the zero range (zr) picture. At $E=0$ the zr-particles move as in a random 
barrier energy landscape. Moves of the zr particles cause changes of the center of mass 
coordinate as defined above: A zr-particle hopping to the right across a bond with 
$s_j>0$ ($s_j<0$) increases (decreases) the center of mass position by $1/L$. 
Conversely, a zr-particle hopping to the left across a bond with $s_j>0$ ($s_j<0$) 
decreases (increases) the center of mass position by $1/L$. As with the field free case 
of the open system, we use a quantum Hamiltonian and a tensor basis for the state space 
to compute the stationary state of the zr lattice of length $M$ with a given 
$\mathbf{s}$ and $\mathbf{w}$. Let $e_1(i)$ denote an unoccupied site $i$ and $e_m(i)$ 
where $m>1$ a site $i$ occupied by $n_i=m-1$ particles. Here, vectors are 
infinite-dimensional. The matrix for creation of a particle at site i is then given by:

\begin{equation}
b_i^{\dagger} = \begin{pmatrix} 0 & 0 & \ldots & 0 & \ldots \\ 1 & 0 & \ldots & 0 & 
\ldots \\ \vdots & \ddots & \ddots & \vdots & \ldots \\ 0 &  & 1 & 0 & \ldots \\ \vdots 
& \ldots & \ldots & \ddots & \ddots \\ \end{pmatrix}_i.
\end{equation}

The matrix for annihilation of a particle at site i is accordingly given by:

\begin{equation}
b_i = \begin{pmatrix} 0 & 1 & \ldots & 0 & \ldots \\ 0 & 0 & 1 & \ldots & \ldots  \\ 
\vdots & \ddots & \ddots & \ddots & \ldots \\ 0 &  & 0 & 0 & \ddots \\ \vdots & \ldots 
& \ldots & \ldots & \ddots \\ \end{pmatrix}_i.
\end{equation}

For constructing the diagonal part of the Hamiltonian we need the following type of 
matrix (note that $ <s|b_i^{\dagger}=<s|{\bf 1} $ and $ <s|b_i=<s|m_i $):

\begin{equation}
m_i= \begin{pmatrix} 0 & 0 & \ldots & \ldots & \ldots \\ 0 & 1 & \ddots &  & \ldots \\ 
\vdots & \ddots & \ddots & 0 & \ldots \\ \vdots  &  & 0 & 1 & \ddots \\ \vdots & \ldots 
& \ldots & \ddots & \ddots \\ \end{pmatrix}_i. 
\end{equation}

A hopping event from site $i$ to site $i+1$ is described by the combined action of the 
matrices $b_i$ and $b_{i+1}^{\dagger}$
This yields the following expression for the quantum Hamiltonian:

\begin{equation}
 H_{zr} = \sum_{i=1}^M \bigg (-h_{i+1}b_{i+1}b_i^{ \dagger}w_{i+1} - h_{i+1}^{-1}
b_{i}b_{i+1}^{ \dagger }w_{i+1} + h_{i+1}^{-1}w_{i+1}m_i + h_{i+1} w_{i+1} m_{i+1} 
\bigg ).
\end{equation}

A yet unnormalized product ansatz 

\begin{equation}
|P_{zr}^*> = \begin{pmatrix} 1 \\ z_1 \\ z_1^2 \\ \vdots \\ \end{pmatrix} \otimes 
\ldots \otimes \begin{pmatrix} 1 \\ z_M \\ z_M^2 \\ \vdots \\ \end{pmatrix}.
\end{equation}

yields 

\begin{eqnarray}
H_{zr}|P_{zr}^*> &  = &  \sum_{i=1}^M \bigg ( -h_{i+1}w_{i+1} \frac{z_{i+1}}{z_i}m_i
|P_{zr}^*> - h_{i+1}^{-1}w_{i+1} \frac{z_{i}}{z_{i+1}}m_{i+1}|P_{zr}^*> \notag \\
& & + h_{i+1}^{-1}w_{i+1}m_i|P_{zr}^*> + h_{i+1} w_{i+1} m_{i+1}|P_{zr}^*> \bigg ) 
\notag \\
&  = &  \sum_{i=1}^M \bigg ( -h_{i+1}w_{i+1} \frac{z_{i+1}}{z_i}m_i|P_{zr}^*> - 
h_i^{-1}w_i \frac{z_{i-1}}{z_i}m_i|P_{zr}^*> \notag \\
& & + h_{i+1}^{-1}w_{i+1}m_i|P_{zr}^*> + h_i w_i m_i|P_{zr}^*> \bigg ).
\end{eqnarray}

$|P_{zr}^*>$ is stationary provided

\begin{equation}
h_{i+1}w_{i+1}z_{i+1} + h_i^{-1}w_iz_{i-1} = h_{i+1}^{-1}w_{i+1}z_i + h_i w_iz_i.
\end{equation}

Applying a general solution \cite{Derrida} to the case at hand and including 
normalization the stationary state for a channel characterized by $\mathbf{s}$ and 
$\mathbf{w}$ and with a total number of particles $K$ yields

\begin{equation}
|P_{zr}^*(\mathbf{s},\mathbf{w},K,M)> =  \sum_{(n_1,...,n_M)}^{\quad \quad ,} 
\prod_{i=1}^M z_i^{n_i} |n_1,...,n_M>  \frac{1}{\sum_{(n_1,...,n_M)}^{,}  
\prod_{i=1}^M z_i^{n_i} },
\end{equation}

where 

\begin{equation}
z_l = \sum_{i=1}^M \frac{1}{h_{l+i}^{-1}w_{l+i}} \prod_{j=1}^{i-1} h_{l+j}^2.
\end{equation}

The primed sums are meant to be summations under the constraint $\sum_{i=1}^{\sigma} 
n_i=K$. Knowing the stationary state for each channel, the drift velocity for the 
individual channels can be computed. In the lattice gas picture, every time a particle 
$s_i=-1$ ($s_i=1$) is hopping to the right (left) it changes the position of the center 
of mass by $1/(L+1)$. The opposite process, i.e. a particle $s_i=-1$ ($s_i=1$) hopping 
to the left (right) changes the center of mass position by $-1/(L+1)$. Therefore, as 
mentioned above, $v=j^--j^+$, i.e. the difference between the currents of negative and 
positive particles. This operator $v$ translates to the zr picture as follows:

\begin{equation}
j^{-}-j^{+} \rightarrow \frac{1}{L} \sum_{i=1}^{M} s_{i+1}w_{i+1}(h_{i+1}^{-1} 
b_{i}b_{i+1}^{\dagger} -  h_{i+1} b_{i+1}b_{i}^{\dagger}).
\end{equation}

Thus, the drift velocity of a given channel is given by 

\begin{eqnarray}
<v(\mathbf{s},\mathbf{w},K,M)> & = & <s|\frac{1}{L} \sum_{i=1}^{M}( s_{i+1}h_{i+1}^{-1}
w_{i+1} b_{i}b_{i+1}^{\dagger} \notag \\
& & - s_{i+1} h_{i+1}w_{i+1} b_{i+1}b_{i}^{\dagger})\Theta (K)| P_{zr}^*>  \frac{1}{c_K}
\end{eqnarray}

Here, $\Theta(K)$ projects on those states of $|P_{zr}^*>$ which have a constant number 
of particles $K$  and $ c_K $ is the normalization. So $\Theta (K) /c_K|P_{zr}^*>=
|P_{zr}^*(\mathbf{s},\mathbf{w},K)> $. Using  this form allows us to make use of the 
fact that due to the combined effect of  $ b_i b^{\dagger}_j $ which is to redistribute 
the particles without changing their number,  we can commute $ b_i b^{\dagger}_j $ and 
$ \Theta (K) $. This allows us then to apply the matrices to the product measure, where 
its effect is easy to see. 

\begin{eqnarray}
<v(\mathbf{s},\mathbf{w},K,M)> & = & <s|\frac{1}{L} \sum_{i=1}^{M} (s_{i+1}h_{i+1}^{-1}
w_{i+1} b_{i}b_{i+1}^{\dagger} \notag \\
& & - s_{i+1} h_{i+1} w_{i+1}b_{i+1}b_{i}^{\dagger}) \Theta (K) |P_{zr}^*>  
\frac{1}{c_K} \notag \\ 
& = & <s|\frac{1}{L} \sum_{i=1}^{M}( s_{i+1}h_{i+1}^{-1}w_{i+1}\Theta (K) b_{i}
b_{i+1}^{\dagger} \notag \\
& & - s_{i+1} h_{i+1} w_{i+1}\Theta (K)b_{i+1}b_{i}^{\dagger})| P_{zr}^*> \frac{1}
{c_K} \notag \\
& = & \frac{1}{L} \sum_i (s_{i+1}h_{i+1}^{-1} w_{i+1} \frac{z_i}{z_{i+1}} <s|m_{i+1}
|\Theta (K)P_{zr}^*> \notag \\ 
& &  - s_{i+1} h_{i+1} w_{i+1} \frac{z_{i+1}}{z_i} <s|m_i |\Theta (K)P_{zr}^*> )  
\frac{1}{c_K}
\end{eqnarray}

We now have to calculate the expression $ c_K $ for the product state: 

\begin{equation}
c_K = \sum_{(n_1,...,n_M)}^{\quad \enspace ,} \prod_{i=1}^M z_i^{n_i} 
\end{equation} 

where the primed sum again means summing with the constraint $ \sum_{i=1}^M n_i = K $.
\newline
A similar type of summation is found when regarding the sums of the type

\begin{equation}
<s|m_{\beta}  \frac{\Theta(K)}{c_K}|P_{zr}^*> = \sum_{(n_1,...,n_M)}^{\quad 
\enspace ,, } \frac{1}{c_K} \prod_{j=1}^{M} z_j^{n_j} ,
\end{equation}

where the double primed sum has the constraints  $ \sum_{j=1}^M n_j = K $ and  
$ n_{\beta} \neq 0 $. 
A straightforward calculation of the sums is impossible due to the constraints. 
To simplify the task, we can profitably use the following identity \cite{Koi}:

\begin{equation}
\sum_{(n_1,...,n_M)}^{\quad \enspace ,} \prod_{i=1}^M z_i^{n_i} = \frac{1}{2 \pi i } 
\oint \frac{d\alpha}{\alpha^{K+1}} \sum_{(n_1,...,n_M)} \prod_{j=1}^{M} 
( \alpha z_j)^{n_j}.
\end{equation}

Thus we can transform the sums into integrals:

\begin{eqnarray}
c_K &  = & \frac{1}{2 \pi i } \oint \frac{d\alpha}{\alpha^{K+1}} \sum_{(n_1,...,n_M)} 
\prod_{j=1}^{M} ( \alpha z_j)^{n_j} \notag \\
& = &  \frac{1}{2 \pi i } \oint \frac{d\alpha}{\alpha^{K+1}} \prod_{j=1}^M \frac{1}{1- 
\alpha  z_j} \notag \\
& = & Q_{K,M},
\end{eqnarray}

\begin{eqnarray}
<s|m_{\beta}  \frac{\Theta(K)}{c_K}|P_{zr}^*> & = & \frac{1}{c_K}  \frac{1}{2 \pi i } 
\oint \frac{d\alpha}{\alpha^{K+1}} \sum_{(n_1,...,n_M),n_{\beta} \neq 0} \prod_{j=1}^{M}
 (\alpha  z_j)^{n_j} \notag \\
& = & \frac{1}{c_K}  z_{\beta}  \frac{1}{2 \pi i } \oint \frac{d\alpha}{\alpha^{K}} 
\prod_{j=1}^M \frac{1}{1- \alpha  z_j} \notag \\
& = & z_{\beta}  \frac{Q_{K-1,M}}{Q_{K,M}}.
\end{eqnarray}

The integrals $Q_{K,M}$ satisfy the recursion relation

\begin{equation}
Q_{K,M}=Q_{K,M-1}+z_M Q_{K-1,M}.
\end{equation}

An explicit solution of this relation yields:

\begin{equation}
Q_{K,M}=\sum_{i=1}^{M} z_i^{K+M-1} \prod_{l=1}^{\enspace M \enspace ,} (z_i -z_l)^{-1}.
\end{equation}

The primed product indicates $i \neq l$.
\newline
From this point on we are not going to carry through the complete calculation for 
$ <v(\mathbf{s},\mathbf{w},K,M)> $. We will expand it into a series in $E$ and keep 
only first order terms which is sufficient for employing the Einstein relation. 
We start from the expression for $ <v(\mathbf{s},\mathbf{w},K,M)> $ after having 
inserted the result for $ <s|m_i  \frac{\Theta (N-L)}{c_K}|P_{zr}^*> $:

\begin{eqnarray}
<v(s,w,K)> & = & \frac{1}{L} \sum_{i=1}^M (s_{i+1}h_{i+1}^{-1} w_{i+1} \frac{z_i}
{z_{i+1}}   z_{i+1}  \frac{Q_{K-1,M}}{Q_{K,M}} \notag \\
& & - s_{i+1} h_{i+1} w_{i+1} \frac{z_{i+1}}{z_i}  z_i  \frac{Q_{K-1,M}}{Q_{K,M}} 
\notag \\
& = & \frac{1}{L} \sum_{i=1}^M s_{i+1}(h_{i+1}^{-1} w_{i+1}  z_i -  h_{i+1} w_{i+1}  
z_{i+1})   \frac{Q_{K-1,M}}{Q_{K,M}} \notag \\
& = & \frac{1}{L} \sum_{i=1}^M s_{i+1}  (1-\exp(-ES))   \frac{Q_{K-1,M}}{Q_{K,M}} 
\notag \\
& = & \frac{S}{L}  (1-\exp(-ES)) \frac{Q_{K-1,M}}{Q_{K,M}},      
\end{eqnarray}

where $S=\sum_{i=1}^Ms_i$. For the second equality we used
 the explicit form of the $z_i$. The term in brackets is easily expanded into a series 
in $E$: $1 - \exp(ES) = E  S + o(E^2)$. We keep only the first order term in $E$. This 
means that when we are expanding the expression $Q_{K-1,M}/Q_{K,M}$ we need only to 
keep zero order terms in $E$, as all other contributions will vanish when using the 
Einstein relation. The only terms containing $E$ in the $ Q_{K,M} $ are the $ z_i $, 
for which we find to first order in $E$:

\begin{eqnarray}
z_i & = &  \sum_{j=1}^{M} \frac{1}{h_{i+j}^{-1}w_{i+j}} \prod_{k=1}^{j-1} h_{i+k}^{2} 
\notag \\
& = & \sum_{j=1}^{M} \frac{1}{w_{i+j}} (1-\frac{E}{2}s_{j+i}+...) \prod_{k=1}^{j-1} 
(1-Es_{i+k}+...) \notag \\
& = & \sum_{j=1}^{M} \frac{1}{w_{i+j}} + o(E). \notag \\
& = & z + o(E).
\end{eqnarray}

Here, $z$ is defined as the zero field value of $z_i$.
The fact that all  $ z_i $ are equal at zero field is decisive for the explicit 
integration of $ Q_{K,M} $. 
We obtain:
\begin{eqnarray}
Q_{K,M} & = &   \frac{1}{2 \pi i } \oint \frac{d\alpha}{\alpha^{K+1}} \prod_{j=1}^M 
\frac{1}{1- \alpha  z_j} \notag \\
& = & { M+K-1 \choose M-1}  z^{K}.
\end{eqnarray}

It is now clear how to proceed with evaluating $ <v(s,w,K,M)> $, as 

\begin{equation}
\frac{Q_{K-1,M}}{Q_{K,M}} = \frac{{K+M-2 \choose M-1} z^{M-1}}{{K+M-1 \choose M-1} 
z^{K}} = \frac{K}{K+M-1}  \frac{1}{z}.
\end{equation}

This yields for $ <v(s,w,K,M)> $:

\begin{eqnarray}
 <v(\mathbf{s},\mathbf{w},K,M)>  & = & \frac{1}{L} \sum_{i=1}^M s_{i+1}  E  S  
\frac{K}{K+M-1}  \frac{1}{z} + o(E^2) \notag \\
& = & \frac{1}{L}  E  S^2  \frac{K}{K+M-1}  \frac{1}{z} + o(E^2).
\end{eqnarray}

This is our final expression for $<v(\mathbf{s},\mathbf{w},K,M)> $.
Knowing the drift velocity for each channel characterized by \textbf{s,w} we have to 
give each of those a weight $\Psi(\mathbf{s,w},K,M)$ when averaging over the channels. 
Following an argument by Pr\"ahofer \cite{Diplom}, we show below that provided a 
configuration in the zr picture is weighted such that at $E=0$ its probability as 
contained in $P_{ring}^*(\mathbf{s,w},K,M)=\Psi(\mathbf{s,w},K,M)P_{zr}^*(\mathbf{s,w},
K,M)$ is equal to the probability of the corresponding state in the lattice gas of the 
open system as given by $P_{open}^*(0)$, for the diffusion constants the following 
relation holds: $D_{L+1}^{open} \geq D_{L}^{per}$. \\
When relating zr to lattice gas probabilities it has to be kept in mind that the 
periodic zr system has $M$ sites, while the corresponding periodic lattice gas has 
$L \geq M$ sites. This means a state of the zr system with probability $q_{zr}$ can be 
permutated $M$ times where due to translational invariance all resulting states have 
the same probability. A lattice gas configuration with probability $q_{lg}$ can be 
cycled through $L$ permutations leading to equal probabilities. Therefore, equal 
probabilities of respective configurations means: $Mq_{zr}=Lq_{lg}$. The weight factor 
thus has to be chosen as follows:

\begin{equation}
\Psi(\mathbf{s,w},K,M) = \frac{d^M}{(2d+1)^L}  {L \choose M}  \prod_{j=1}^M f(w_j).
\end{equation}

Thus we find for the drift velocity when averaged over the channels:

\begin{equation}
\begin{split}
\bar{v} & =  \sum_M \sum_{\mathbf{s}=(s_1,...,s_M)} \sum_{\mathbf{w}=(w_1,...,w_M)} 
\Psi(\mathbf{s},\mathbf{w})  <v(\mathbf{s},\mathbf{w},K,M)> \\
& = \sum_M \frac{1}{L}  E  M   \frac{d^M}{(2d+1)^L}  { L \choose M }   \frac{L-M}{L-1} 
 \left(\sum_{w} \prod_{j=1}^M f(w_j) \frac{1}{z} \right). 
\end{split}
\end{equation}

This can be rewritten as 

\begin{equation}
 \overline{v} = E  \sum_{M=1}^L \Omega(M)  M  \big < \frac{1}{z} \big >. 
\end{equation}

Here $\Omega(M)$ contains all the factors depending on $M$ occurring in the previous 
expression and $<1/z>$ is a disorder average. $\Omega(M)$ is a function that is sharply 
peaked at $M=L/2$ implying that in the limit for large $L$ only terms with large $M$ 
significantly contribute to the result. In the limiting case of $L \to \infty$ invoking 
the central limit theorem (remember that we demanded $<1/W>$ and $<1/W^2>$ to be finite)
 yields:

\begin{equation}
\bigg \langle \frac{1}{z} \bigg \rangle = \frac{1}{M \langle 1/W \rangle}.
\end{equation}

Therefore

\begin{equation}
\overline{v}  =  \sum_{M=1}^L \frac{1}{L}  E  M   \frac{d^M}{(2d+1)^L}  { L \choose M }
   \frac{L-M}{L-1}  \frac{1}{M}  \frac{1}{ \big < \frac{1}{W} \big > }.
\end{equation}

Using the Einstein relation yields in the limit for large $L$ 

\begin{eqnarray}
D(0) & = & \frac{1}{L(L-1)}  \frac{1}{(2d+1)}  (1-\frac{1}{(2d+1)^{L-1}})  \frac{1}{ 
\big < \frac{1}{W} \big > } \notag \\
& = & D^*  \frac{1}{ \big < \frac{1}{W} \big > },
\end{eqnarray}

where $D^*$ is the diffusion constant for the ordered case \cite{Koi,Prae}. \\
Thus we have shown that $\lim_{L \to \infty} DL^2=1/[(2d+1)<1/W>]$ for the periodic 
case, which is a lower bound to the open case. Note that the factor $1/<1/W>$ is the 
same as occurring in the single particle diffusion constant for the random barrier model.
 In this model random energy barriers are assigned to bonds between sites just as in the
 zr picture of our problem.

\section{Model with open boundary conditions}

\subsection{Variational formula}

In order to find an upper bound on the diffusion constant we follow the strategy of Ref. 
\onlinecite{Prae}.  The quantum Hamiltonian $H_{open}$ for our model in its 
representation using the tensor basis can be decomposed into the sum of a diagonal part 
($D$) and a non diagonal one ($M$). Each of these can themselves be split into a part 
describing those moves leading to an increase of the center of mass coordinate ($M^+$ 
and $D^+$) and a respective part connected to a decrease ($M^-$ and $D^-$). Thus 
$H=D^++D^--M^+-M^-$. \\
The following statement \cite{Spohn} holds for the diffusion constant $D$:

\begin{equation}
\begin{split}
D & = \inf_{g \in \Omega} \quad \bigg [ \frac{1}{2(L+1)^2} <D^++D^->-\frac{2}{(L+1)} 
<s|[D^+-D^-]g|P^*>+<s|gHg|P^*> \bigg ] \\
& = \inf_{g \in \Omega}  F[g]. 
\end{split}
\end{equation}

Here, $\Omega$ is the space of diagonal matrices with dimension $(2 \sigma +1)^L$. 
Thus plugging in any diagonal matrix $g$ into the functional $F[g]$ yields an upper 
bound on $D$. The challenge is to choose $g$ such that the upper bound gets as small as 
possible. Due to the dimensionality of the diagonal matrices increasing exponentially 
with $L$, an exact minimization is not feasible. Still, some information can be gained 
by observing that the functional is convex and thus the minimizing $g_0 \in \Omega$ is 
the only matrix for which $\delta F[g]$ vanishes. Choosing $g_0$ such that

\begin{equation}
<s|g_0H=<s|[ D^+-D^-]
\end{equation}

the variation vanishes. Unfortunately, this formula cannot be solved for $g_0$. We will 
show below how still some insight can be gained from this equation.
Introducing the matrix $d$, where $d(y'|y)=<y'|d|y>$ gives the change in the CMS 
coordinate when a transition from state $y$ to $y'$ is made and the matrix $w$, where 
$w(y'|y)$ denotes the corresponding rate, the variational formula may be written as:

\begin{equation}
D= \quad \text{inf} \quad \frac{1}{2} \sum_{y',y \in X} w(y'|y) P^*(y)[d(y'|y)+g(y')
-g(y)]^2,
\end{equation}

where $g(y)=<s|g|y>$.
Following Ref. \onlinecite{Diplom} we now prove that $D^{open}_{L} \geq D^{per}_{L+1}$. 
To simplify the notation we remark that in our model with periodic boundary conditions 
as in the original RD model transitions from a given state $y$ are only allowed to a 
state $y'=y^{i,i+1}$, where the spins $y_i$ and $y_{i+1}$ are interchanged. Thus the 
last formula applied to the periodic case reads as

\begin{equation}
D_{L+1}^{per}= \quad \text{inf} \quad \frac{1}{2} \sum_{l=0}^L \left ( \sum_{y \in X} 
w(y^{l,l+1}|y) P^*(y)[d(y^{l,l+1}|y)+g(y^{l,l+1})-g(y)]^2 \right ).
\end{equation}

Here, each transition between states in a lattice gas of length $L+1$ changes the cm 
coordinate by $\pm 1/(L+1)$, as there are as many reptons as bonds between them. In 
contrast to that in the open boundary case there is one repton more as there are bonds 
leading to the cm coordinate changing by $\pm 1/(L+1)$ in any transition of a lattice 
gas of length $L$. Here, transitions between states can not only occur by interchanging 
spins $y_1$ and $y_{i+1}$ in the bulk, but also by creation and annihilation events at 
the ends. These events can equivalently be viewed as interchanges of the spins $y_1$ or 
$y_L$ respectively with 'imaginary' spins $y_0$ and $y_{L+1}$ provided the rates for 
these interchanges at the ends are adapted such that they correspond to the creation and
 annihilation rates as demanded by the model. Taking the distribution at the 'imaginary'
 sites as for any other site in the lattice gas (remember that we have a homogeneous 
product measure) the introduction of a parameter

\begin{equation}
c_l =
\begin{cases}
2d+1 & \text{if} \quad l=0 \quad \text{or} \quad l=L \\
1 & \text{for} \quad l=1,..,L-1 
\end{cases}
\end{equation}

leads to a fulfillment of this requirement and the functional for the diffusion constant
 of the open chain of length $L$ reads as

\begin{equation}
D_L^{open}= \text{inf} \quad \frac{1}{2} \sum_{l=0}^L c_l \left ( \sum_{y \in X} 
w(y^{l,l+1}|y) P^*(y)[d(y^{l,l+1}|y)+g(y^{l,l+1})-g(y)]^2 \right )
\end{equation}

where $w$ and $d$ are as for the periodic case. Letting $\mathbf{y}=(y_0,..,y_{L+1})$ 
for open and periodic system and observing that as $y_0$ does not occur in the 
functional for $D_{L+1}^{per}$ the averaging of $y_0$ over $P^*$ yields 1 and thus does 
not change the result.

\begin{eqnarray} 
& & D_L^{open}[g]-D_{L+1}^{per}[g] \notag \\
& = & c_0 \left( \sum_{y \in X} w(y^{0,1}|y) P^*(y)[d(y^{0,1}|y)+g(y^{0,1})-g(y)]^2 
\right) \notag \\
& & + c_L \left( \sum_{y \in X} w(y^{L,L+1}|y) P^*(y)[d(y^{L,L+1}|y)+g(y^{L,L+1})-
g(y)]^2 \right) \notag\\
& & - \sum_{y \in X} w(y^{L+1,1}|y) P^*(y)[d(y^{0,1}|y)+g(y^{0,1})-g(y)]^2 \notag \\
& & - \sum_{y \in X} w(y^{L,L+1}|y) P^*(y)[d(y^{L,L+1}|y)+g(y^{L,L+1})-g(y)]^2 \notag \\
& = & 2 \sum_{y \in X} w(y^{L+1,1}|y) P^*(y)[d(y^{0,1}|y)+g(y^{0,1})-g(y)]^2 \notag \\
& & + 2 \sum_{y \in X} w(y^{L,L+1}|y) P^*(y)[d(y^{L,L+1}|y)+g(y^{L,L+1})-g(y)]^2 
\notag \\
& & \geq 0
\end{eqnarray}

Therefore, the diffusion constant of the periodic chain with length $L+1$ is a lower 
bound for the open chain with length $L$.

\subsection{Functional for diffusion with kinematic disorder}

The space of diagonal matrices $\Omega$ from which an appropriate $g \in \Omega$ has to 
be chosen such that $F[g]$ gets as small as possible, has dimension $(2 \sigma+1)^L$. A 
scalar product on $\Omega$ can be defined for arbitrary $m,n  \in \Omega$ as 
$<s|mn|P^*>$. A basis for $\Omega$ can be built from the matrices $\hat{y}_1(i)$ to 
$\hat{y}_{2 \sigma}(i)$ acting non trivially only at site $i$ and the unit matrix. We 
specify only the following:

\begin{equation}
\hat{y}_{\alpha}(i)= \frac{1}{2d+1} E_{(2 \alpha,2 \alpha)}(i)-\frac{1}{2d+1} 
E_{(2 \alpha+1,2 \alpha+1)}(i) \text{ for } \alpha = 1,..,\sigma,
\end{equation}

\begin{equation}
\hat{y}_{\sigma+1}(i)= - \frac{2d}{2d+1} E_{(1,1)}(i) + \sum_{j=2}^{2 \sigma +1} 
\frac{1}{2 d+1} E_{(j,j)}(i).
\end{equation} 

For the remaining $ \sigma -1 $ matrices we demand that they are chosen such that for 
all $ a,b = 1,..,2 \sigma $ , $i,j = 1,..,L$ and $ a \neq b $ we have 
$<\hat{y}_{a}(i)\hat{y}_{b}(j)>= 0$ as well as $<\hat{y}_{a}(i)>=0$. It is easily 
checked that the defined matrices are mutually orthogonal. A basis for the space of 
diagonal matrices $\Omega$ is then given by:

\begin{equation}
\left \{ \prod_{j=1}^{2 \sigma +1} \hat{y}_j(I_j)|I_j \subset \{ 1,..,L \} \quad \forall
 j; I_{\alpha} \cap I_{\beta} = \varnothing \quad \forall \alpha \neq \beta \right \}.
\end{equation}

The choice of a $g \in \Omega$ fulfilling $<s|[D^+-D^-]=<s|gH$ leads to $\delta F[g]=0$.
 In spite of this relation not being solvable for $g$ information can be gained by 
observing that $\Omega$ contains subsets $\Omega_{i_1,..,i_{\sigma}}$ which are 
invariant to $H$ in the sense that if $g \in \Omega_{i_1,..,i_{\sigma}}$, then the 
diagonal matrix $g'$ fulfilling $<s|gH=<s|g'$ is in $\Omega_{i_1,..,i_{\sigma}}$ too, 
where

\begin{equation}
\Omega_{i_1,..,i_{\sigma}}=\text{span} \bigg \{ \prod_{j=1}^{2 \sigma+1} \hat{y}_j(I_j)|
 I_j \subset \{1,..,\} \quad \forall j; I_{\alpha} \cap I_{\beta}=\varnothing \quad 
\forall \alpha \neq \beta; |I_j|=i_j \quad \forall j=1,.., \sigma \bigg \}.
\end{equation}

By calculating $<s|[D^+-D^-]$ we find that the minimizing $g_0$ is in the subspace 
$ \bigoplus_{j=1}^{\sigma} \Omega_{i_1,..,i_{\sigma}}$ where $i_k= \delta_{i,k} \quad 
\forall \quad k=1,..,\sigma$. Thus the most general ansatz for $g$ based on this 
information is:

\begin{equation}
g=\sum_{\alpha=1}^{\sigma}\sum_{i=0}^{L+1}\sum_{I_{\sigma+1},..,I_{2 \sigma+1}}
c_{i,I_{\sigma+1},..,I_{2 \sigma+1}}^{\alpha} \hat{y}_{\alpha}(i)\hat{y}_{\sigma+1}
(I_{\sigma+1})..\hat{y}_{2 \sigma+1}(I_{2 \sigma+1})
\end{equation}

where the coefficients $c$ are real numbers and the sets $I_{\sigma},..,I_{2 \sigma+1}$ 
are mutually disjoint.
\subsection{Ansatz for $g$}
The general ansatz given above for $g$ contains too many parameters for letting a 
minimization appear feasible. Instead we use a generalization of the ansatz used by 
Pr\"ahofer \cite{Diplom} for the ordered RD model:

\begin{equation}
\begin{split}
g= \sum_{ \alpha=1 }^{ \sigma } \bigg \{ & \sum_{i=0}^{L+1}a_i^{\alpha} \hat{y}_{ 
\alpha} (i) +\sum_{k,k'=0}^{L+1 \enspace ,}c_{k,k'}^{ \alpha } \hat{y}_{ \alpha} (k) \hat{y}_
{ \sigma +1} (k') \\
& +\sum_{o,p,q=0;p<q}^{L+1 \enspace ,} e_{o,p,q}^{ \alpha} \hat{y}_{ \alpha}(o)\hat{y}_{ 
\sigma +1}(p)\hat{y}_{ \sigma +1}(q) \bigg \}.
\end{split}
\end{equation}

The primed sums indicate the summation variables to be mutually different. Note that 
this ansatz reflects a particular choice of parameters for the general ansatz. Inserting
 the trial function into $F[g]$ yields the following functional:

\begin{equation}
\begin{split}
& D[g] \\
& = \sum_{ \alpha=1 }^{ \sigma } \bigg \{ \frac{2 d}{(2 d+1)^2} f(W_{ \alpha}) 
W_{ \alpha} \sum_{l=0}^{L} c_l  \\
& \quad  \bigg [ \bigg ( \frac{1}{L+1} +(a_l^{ \alpha}-a_{l+1}^{ \alpha})-\frac{2 d}
{(2 d+1)}(c_{l,l+1}^{ \alpha}-c_{l+1,l}^{ \alpha}) \bigg )^2 \\
& \quad +\frac{2 d}{(2 d+1)^2}    \sum_{k'}" \bigg (c_{l,k'}^{ \alpha}-{c_{l+1,k'}^{ 
\alpha}}- \frac{2 d}{(2 d+1)} (e_{l,(l+1,k')}^{ \alpha}-e_{l+1,(l,k')}^{ \alpha}) 
\bigg )^2 \\
& \quad + \sum_{ \beta=1}^{ \sigma} \frac{2 d}{(2 d+1)}f(W_{\beta}) \sum_{k}"
(c_{k,l}^{ \beta}-{c_{k,l+1}^{ \beta}})^2 \\
& \quad + \frac{4d^2}{(2 d+1)^4 } \sum_{p<q}"(e_{l,p,q}^{ \alpha}-e_{l+1,p,q}^{ \alpha}
)^2 \\
& \quad + \sum_{ \beta=1}^{ \sigma} \frac{4d^2}{(2 d+1)^3 }f(W_{ \beta}) \sum_{o,q}
(e_{o,(l,q)}^{ \beta}-e_{o,(l+1,q)}^{ \beta})^2 \bigg ] \bigg \}. 
\end{split}
\end{equation}

The free parameters are the coefficients $a_i^{\alpha},c_{k,k'}^{\alpha}$ and 
$e_{o,p,q}^{\alpha}$ which are, generalizing results in Ref. \onlinecite{Prae} 
chosen as

\begin{equation}
\begin{split}
& a_{i+1}^{ \alpha}=\sum_{l=0}^i \bigg ( \frac{1}{L+1} -\frac{2 d}{(2 d+1)}
(c_{l,l+1}^{ \alpha}-c_{l+1,l}^{ \alpha}) \bigg ) \\
& a^{ \alpha}_0=0
\end{split}
\end{equation}

\begin{equation}
c_{i,j}^{ \alpha}= 
\begin{cases}
\frac{(2 d+1)C}{2 d}(1-g(\frac{i}{n}))\frac{L+1-j}{(L+1)^2} & \text{ for } i<j \\
-c_{L+1-i,L+1-j}^{\alpha} & \text{ for } i>j \\
\end{cases}
\end{equation}

\begin{equation}
e_{o,p,q}^{ \alpha}=
\begin{cases}
\frac{(2 d+1)^2 C}{4 d n}g'(\frac{o}{n})\frac{(L+1-p)(L+1-q)}{(L+1)^2(L-o)} & 
\text{ for } o<p<q \\
-e_{L+1-o,L+1-q,L+1-p}^{\alpha} & \text{ for } p<q<o. \\
\end{cases}
\end{equation}

Here, $g$ is a monotonically decreasing smooth function on the real numbers with 
$g(x)=1$ for $x \leq 0$ which is decreasing exponentially fast for 
$x \rightarrow \infty$. This function has the property that

\begin{equation}
\sum_{i=0}^{\infty}\left[g^{(r)} \bigg ( \frac{i}{n} \bigg ) \right]^s=O(n),
\end{equation}

where $r,s$ are integers, and $n=(L+1)^{0.75}$. Given this choice, the following upper 
bound for $D_{L+1}^{open}$ is found

\begin{equation}
D_{L+1}^{open} \leq \frac{1}{(2 d+1)} \frac{1}{(L+1)^2} < W >,
\end{equation}

where $<W>$ is the disorder average.

This completes the proof that asymptotically $D \propto 1/L^2$ in the presence of 
kinematic disorder.

\section{MC results}

\begin{figure}[h]
\begin{center}{ \epsfig{figure=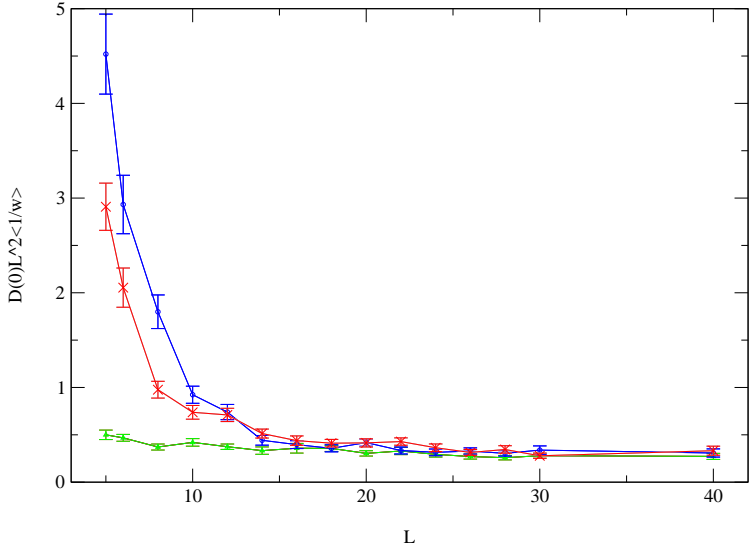,width=9cm, angle=-90}}
\caption{$D(0)L^2<1/W>$ versus $L$ for the cases as described in the text.}
\label{vergleich}
\end{center}
\end{figure}

In the case of the ordered RD model the results for $D$ in the periodic case \cite{Koi} 
coincide to leading order with the upper bound for the open case \cite{Prae}. In our 
model we managed to give bounds on $D$ which both scale with $1/L^2$ in the limit for 
long chains. This implies that kinematic disorder does not ruin the scaling relation of 
standard reptation theory. Still, the numerical prefactors differ. We believe that it 
is the result for the lower bound which correctly describes the diffusive behavior for 
long chains. This lower bound was obtained by a rigorous calculation while the upper 
bound results from a variational treatment. Given that our conjecture is true and for 
long chains $D=1/[L^2(2d+1)<1/W>]$ then for any choice of disorder distribution 
$DL^2<1/W>$ plotted against the chain length is constant. We performed MC simulations 
with different disorder distributions. It turns out that $L=30$ is sufficient to 
observe the curves all converging to the same constant value. For the cases $<1/W>=1$,
 $<1/W>=0.02$ and $<1/W>=0.03$ figure \ref{vergleich} shows our MC results. These results suggest 
that in the long chain length limit $D=1/[L^2(2d+1)<1/W>]$.

\section{Dynamics of internal segments}

The surprising result of the previous sections is the observation that the effect of 
kinematic disorder on the collective behavior of all connected  polymer segments is 
(to leading order in system  size) the same as on a simple point like object in the 
same disorder environment. In order to understand this observation we now consider
the dynamics of the internal segments in the hydrodynamic limit of vanishing
lattice spacing. For the local concentration of reptons one obtains
from the usual ordered repton model in this limit Rouse dynamics \cite{Exp2}
restricted to motion inside the tube \cite{Doi}. The boundary dynamics, i.e.,
the hopping into and out of the tube resp. describe the entropic tensile
force acting on the chain ends and keeping the polymer in its stretched
equilibrium conformation. For understanding the dependendence of the diffusion
coefficient on kinematic disorder, which affects mostly the bulk of the
polymer chain, it is sufficient to focus on the hydrodynamic behavior of the
bulk reptons. In order to study this limit it is convenient to first investigate the
associated zero-range process and then translate the result into the
hydrodynamic limit of the exclusion process.

For the zero-range process described in Section III with $E=0$, the
average number of particles $\rho_i(t)$ at site $i$ obeys the exact
time evolution equation given by  
\begin{equation}
\frac{\partial \rho_{i}(t)}{\partial t}=W_{i} \; z_{i-1}(t) + W_{i+1} \;
z_{i+1}(t)- (W_{i} + W_{i+1}) \; z_{i}(t) \;\;\;\; \label{extime}
\end{equation}
where $z_{i}(t)$ is the probability that site $i$ is occupied at time
$t$. It is known that in the steady state, the occupancy probability
$z_{i}^{ss}$ is spatially uniform and can be related to the steady state
particle density as \cite{Richard}
\begin{equation}
z_{i}^{ss}=\frac{\rho_{i}^{ss}}{1+ \rho_{i}^{ss}} \label{srho}.
\end{equation}
It follows that the steady state density profile is uniform in spite
of disorder. 

To understand the dynamics in the hydrodynamic limit of vanishing
lattice spacing, we expand Eq.(\ref{extime}) to second order in
lattice constant and find  
\begin{equation}
\frac{\partial \rho(x,t)}{\partial t}=\frac{\partial}{\partial x}
\left[ {W(x) \; 
\frac{\partial z(x,t)}{\partial x}} \right] \;\;\;\;.
\end{equation}
At large enough time, the system is expected to be in local
equilibrium so that we may assume the usual approximation for
$z(x,t)-\rho(x,t)$ in the steady state to be valid:
$z(x,t) \approx \rho(x,t)/(1+\rho(x,t))$, where equation (\ref{srho})
was used. 
We are interested in the density fluctuations about the steady state,
$\Delta \rho(x,t)=\rho(x,t)-\rho^{ss}(x)$ where
$\rho^{ss}(x)=\rho=K/M$. Retaining the 
lowest nonvanishing term in the expansion of the preceding equation in
powers of $\Delta \rho(x,t)$, we obtain
\begin{equation}
\frac{\partial \Delta \rho(x,t)}{\partial t}=\frac{1}{(1+\rho)^{2}}
\;\frac{\partial}{\partial x} \left[ {W(x) \; \frac{\partial \Delta
\rho(x,t)}{\partial x}} \right]  \label{drho} \;\;\;\;.
\end{equation} 
The above equation describes a random walker in one dimension,
diffusing in a random medium with 
bond-symmetric hopping rates $W(x)$. It can be shown that at large time and
length scales, the random walker can be described by a single,
effective diffusion constant, ${\cal{D}}=1/\left[ (1+\rho)^{2}
\;\langle {1/W} \rangle \right]$, provided $\langle {1/W} \rangle$ is finite \cite{bondsymmRW}. Thus we
obtain the bulk diffusion constant to be given by ${\cal{D}}$ in the
zero range process.  

Regarding the site as particle and mass as hole clusters, the above
model maps onto symmetric exclusion process (SEP) with particlewise disorder.
We want to calculate the bulk diffusion constant in the SEP picture
using the above results for zero range process. Since the steady state
density profile is uniform in both the pictures, the average
local density $n_{i}$ in the vicinity of the location of particle $i$ in SEP 
is related to that in the zero range process as 
$n_{i}=1/(1+\rho_{i})$. Then the density fluctuation about the steady
state in the vicinity of particle $i$ is given by 
\begin{equation}
\Delta n_{i} = \frac{-1}{(1+\rho)^2} \Delta \rho_{i} +O((\Delta \rho)^{2})
\;\;\;\;.\label{nrho} 
\end{equation}
We note that the above is true only at large enough times, as was
pointed out in a similar analysis for the tagged particle correlation
function for SEP without disorder \cite{alexpinc}.

Using Eq.(\ref{drho}) and Eq.(\ref{nrho}), we find that
\begin{equation}
\frac{\partial \Delta n(x,t)}{\partial t}=\frac{1}{(1+\rho)^{2}}
\;\frac{\partial}{\partial x} \left[ {W(x) \; \frac{\partial \Delta
n(x,t)}{\partial x}} \right] \;\;\;\;.
\end{equation}
Further note that $x$ is the space index in the zero range process,
while it labels the particles in the SEP. The space coordinate $y$ in
symmetric exclusion process is related to $x$ as
\begin{equation}
y \approx \int^{x} dx^{\prime} \rho(x^{\prime}) + x  \;\;\;\;,
\end{equation}
which gives
\begin{equation}
\frac{\partial}{\partial x}=(1+\rho) \frac{\partial}{\partial
y}+O(\Delta n)  \;\;\;\;.
\end{equation}
Thus the particle density fluctuations in the SEP obey
\begin{equation}
\frac{\partial \Delta n(y,t)}{\partial t}= \frac{\partial}{\partial y}
\left[ {W(y) \; \frac{\partial \Delta n(y,t)}{\partial y}}\right]  \;\;\;\;.
\end{equation}
Using the random walker analogy, we obtain the
effective diffusion constant $\overline{{\cal{D}}}$ at large times to
be equal to $1/\langle {1/W} \rangle$ in the symmetric exclusion process. 

This calculation shows that the internal segments of the polymer chain 
perform Rouse-type dynamics also in the presence of kinematic disorder, but with a 
disorder-dependent diffusion coefficient. 
This explains the occurrence of the
same correction to the diffusion coefficient for the long time behavior
of the polymer chain as a whole.

\section{Conclusions}

It is the aim of this paper to disentangle the effects of the various
types of disorder which one may expect to have significant impact on the
dynamics of systems of entangled flexible polymers. We have focused
on kinematic disorder which leaves the equilibrium conformation unchanged
compared to a hypothetical ordered entanglement network (which could in
principle be manufactured artificially by placing a single polymer
in an ordered array of obstacles on a surface). We have proved (rigorously
in terms of the RD-model for reptation) that the asymptotic length scaling of the
diffusion coefficient of the polymer chain remains as predicted by
standard reptation theory. By studying the hydrodynamic limit we have shown
that the individual polymer segments inside the tube
perform Rouse dynamics in a disordered environment which corresponds to a
system of local random barriers. Therefore the amplitude of the diffusion
coefficient becomes dependent on the disorder in the same way a single particle
in a random barrier system.

\section{Acknowledgment}

K.J. wishes to thank the Forschungszentrum J\"ulich for kind hospitality.
R.D.W. thanks Deutsche Forschungsgemeinschaft for financial support.

\end{document}